\documentclass{LMCS}
\usepackage{xspace}
\usepackage{amssymb}
\usepackage{enumerate}
\usepackage{hyperref}

\theoremstyle{plain}\newtheorem{claim}[thm]{Claim}

\newcommand{\cD}{{\ensuremath {\mathcal D}}}
\newcommand{\cN}{{\ensuremath {\mathcal N}}}
\newcommand{\cS}{{\ensuremath {\mathcal S}}}
\newcommand{\cR}{{\ensuremath {\mathcal R}}}

\newcommand{\comment}[1]{}
\newcommand{\zug}[1]{\ensuremath{\langle {#1} \rangle}}
\renewcommand{\inf}{\mathit{inf}}
\newcommand{\shorten}[2]{#1}
\newcommand{\myproof}[2]{{\begin{proof}{#2}\end{proof}}}
\newcommand{\Secl}[1]{\label{Section:{#1}}}
\newcommand{\Secr}[1]{Section~\ref{Section:{#1}}}
\newcommand{\Thml}[1]{\label{Theorem:{#1}}}

\newcommand{\Clml}[1]{\label{Claim:{#1}}}
\newcommand{\Clmr}[1]{Claim~\ref{Claim:{#1}}}

\newcommand{\safrastates}{(12)^{n}n^{2n}}
\newcommand{\safrarstates}{(12)^{n(k+1)}(n(k{+}1))^{2n(k+1)}}
\newcommand{\safrasstates}{(12)^{n(k+1)}n^n(k{+}1)^{n(k+1)}(n(k{+}1))^{n(k+1)}}
\newcommand{\dpwsstates}{2n^n(k{+}1)^{n(k+1)}(n(k{+}1))!}

\def\doi{3 (3:5) 2007}
\lmcsheading%
{\doi}
{1--21}
{}
{}
{Nov.~\phantom{0}9, 2006}
{Aug.~14, 2007}
{}   

\begin{document}

\title[Determinization of B\"uchi and Streett Automata]{From Nondeterministic B\"uchi and Streett Automata to Deterministic Parity Automata\rsuper*}

\author[N.~Piterman]{Nir Piterman}	%required
\address{Ecole Polytechnique F\'ed\'eral de Lausanne (EPFL)}	%required
\email{nir.piterman@epfl.ch}
%\thanks{thanks 1, optional.}	%optional

%% etc.

%% required for running head on odd and even pages, use suitable
%% abbreviations in case of long titles and many authors:

%% mandatory lists of keywords and classifications:
\titlecomment{{\lsuper*}This is an extended version of \cite{Pit06}.}
%%%%%%%%%%%%%%%%%%%%%%%%%%%%%%%%%%%%%%%%%%%%%%%%%%%%%%%%%%%%%%%%%%%%%%%%%%%

%% the abstract has to PRECEED the command \maketitle:
%% be sure not to issue the \maketitle command twice!

\begin{abstract}
In this paper we revisit Safra's determinization constructions for
automata on infinite words.
We show how to construct deterministic automata with fewer states
and, most importantly, parity acceptance conditions.
Determinization is used in numerous applications, such as reasoning
about tree automata, satisfiability of CTL$^*$, and
realizability and synthesis of logical specifications.
The upper bounds for all these applications are reduced by using the
smaller deterministic automata produced by our construction.
In addition, the parity acceptance conditions allows to use more
efficient algorithms (when compared to handling Rabin or Streett
acceptance conditions).
\end{abstract}

\maketitle

%% start the paper here:
\section{Introduction}
\Secl{intro}
One of the fundamental questions in the theory of automata is 
determinism vs. nondeterminism.
Another related question is the question of complementation. 
That is, given some machine in some complexity class can we produce
a machine in the same class that accepts the complement language?
The problems of determinization and complementation are strongly
related.
Indeed, if the machine is deterministic we just have to dualize its
answer.  
If the machine is nondeterministic we do not have a simple solution.

% unclear sentence - orna
In the theory of finite automata on finite words the relation between
nondeterministic and deterministic automata is well understood.
%In the theory of finite automata on finite words these two
%questions are resolved. 
We know that there exists an efficient procedure that gets a
nondeterministic automaton with $n$ states and constructs a
deterministic automaton with $2^n$ states accepting the same language
\cite{RS59}. This construction is tight \cite{HMU00}.
By dualizing the acceptance condition of the deterministic automaton
we get an automaton for the complement language, which is again tight
\cite{HMU00}.

% See comment 15 - orna
In his proof that satisfiability of S1S is decidable, B\"uchi
introduces nondeterministic automata on infinite words
\cite{Buc62}.
B\"uchi takes a `normal' finite automaton and runs it on infinite
words. 
A run of such an automaton is an infinite sequence of states, instead
of a finite sequence.
The set of states {\em recurring} infinitely often is used to
define the acceptance condition. 
A run is accepting according to the {\em B\"uchi condition} if the set
of recurring states intersects the set of accepting states.

In the case of finite automata on infinite words determinization and
complementation are much more involved.  
Given a deterministic B\"uchi automaton one can easily construct a
nondeterministic B\"uchi automaton for the complement language
\cite{Kur87}.
However, deterministic B\"uchi automata are not closed under
complementation \cite{Lan69}. 
This forced the introduction of more complex acceptance conditions
such as Muller, Rabin, Streett, and parity.
A Rabin acceptance condition is a set of pairs of subsets of
the states. 
A run is accepting according to a Rabin condition if there exists a
pair $\zug{E,F}$ such that the set of recurring states does not
intersect $E$ but does intersect $F$. 
The Streett condition is the dual of Rabin.
A run is accepting according to a Streett condition if for every
pair $\zug{R,G}$ we have that if $G$ intersects the set of recurring
states so must $R$.
A parity condition gives an integer priority to every state and a run
is accepting if the minimal recurring priority is even.
The number of priorities is the {\em index} of the parity condition.
Rabin and Streett conditions are more general than parity in the
following sense.
A parity condition of index $2k$ can be written as a Rabin (or
Streett) condition with $k$ pairs (without modifying the structure of
the automaton).
We can translate a Rabin or Streett condition with $k$ pairs to a
parity condition of index $2k+1$ using a gadget with $k^2k!$ states,
thus we multiply the number of states of the automaton by $k^2k!$.
All three conditions are strong enough to allow determinization
\cite{Tho90}.

In the case of automata on infinite words determinization and
complementation are no longer so strongly coupled.
% not clear - orna
Determinization can be used for complementation by dualizing the
acceptance condition of the deterministic automaton.
However, there are complementation constructions that are much simpler
than determinization.
%It is still the case that determinization allows complementation but
%the latter may be simpler if it avoids the intermediate
%determinization.
Specifically, B\"uchi showed that the class of languages recognized by
nondeterministic B\"uchi automata is closed under complement without
determinization \cite{Buc62}.
Sistla, Vardi, and Wolper suggested a singly exponential
complementation construction \cite{SVW85}, however with a quadratic
exponent. 
This was followed by a complementation construction by
Klarlund \cite{Kla91} and a very elegant complementation via alternating
automata by Kupferman and Vardi \cite{KV01c}. 
%This was followed by a very elegant complementation
%construction of Kupferman and Vardi \cite{KV01c}.
The latter construction was 
recently improved to give a complement automaton
with at most $(0.96n)^n$ states \cite{FKV04}, 
which is currently the best complementation construction. 
\shorten{See also \cite{Tho90}.}{}

Determinization constructions for automata on infinite words followed
a similar path\footnote{%
  Incidentally, both determinization constructions provided the
  best upper bound for complementation at the time of their
  introduction.%
}.
McNaughton showed a determinization construction that is doubly
exponential and results in an automaton with the Muller acceptance
condition \cite{Mcn66}. 
Safra gives a determinization construction which takes a
nondeterministic B\"uchi automaton with $n$ states and returns a
deterministic Rabin automaton with at most $\safrastates$ states and $n$
pairs \cite{Saf88}.
An alternative determinization with a similar upper bound that also
results in a deterministic Rabin automaton was given by Muller and
Schupp \cite{MS95}.
Michel showed that this is asymptotically optimal and that the best
possible upper bound for determinization and complementation is $n!$
\cite{Mic88,Lod98}. 

% Explain Safra - orna
Safra's idea is to use multiple subsets for one state of the
deterministic automaton and organize them in the form of a tree.
The root of the tree is the classical subset construction for automata
on finite words.
In every transition, a node with set of states $S$ spawns a new son
that includes all the accepting states in $S$.
Thus, all the states in a leaf are the endpoints of runs that agree
(more or less) on the number of times they have visited the
acceptance set.
In order to keep the tree finite, we ensure that every state is
followed in at most one branch of the tree.
Whenever a state is followed in more than one branch
we keep only the copy in the oldest branch.
Furthermore, whenever all the states followed by some node have
visited the acceptance set, the node is marked as accepting and all
its descendants are removed.
The Rabin acceptance condition associates a pair with every 
node in the tree. 
There should be some node that is erased from the tree at most
finitely often and marked accepting infinitely often for a run to be
accepting.

The fact that stronger acceptance conditions are introduced raises
the question of determinization of automata using these conditions.
Rabin and parity automata can be easily converted to B\"uchi automata.
%Thus, determinization of B\"uchi automata solves determinization for
%Rabin and parity automata.
Given a Rabin automaton with $n$ states and $k$ pairs there exists an
equivalent nondeterministic B\"uchi automaton with $n(k+1)$ states.
Applying Safra's determinization on top of this automaton produces a
deterministic Rabin automaton with $\safrarstates$ states and $n(k+1)$
pairs.
%A parity condition of index $k$ can be expressed by a Rabin condition
%with $\frac{k}{2}$ pairs. 
%Thus, determinization of parity automata can be solved like
%determinization of Rabin automata.
For Streett automata, going through nondeterministic B\"uchi automata
is far from optimal.
A nondeterministic Streett automaton with $n$ states and $k$ pairs can
be converted to a nondeterministic B\"uchi automaton with $n2^k$
states \cite{Cho74}, which is optimal \cite{SV89}.
Combining this conversion with the determinization results in a
doubly exponential deterministic automaton.
In order to handle Streett automata, Safra generalized his
determinization construction \cite{Saf92}. 
Given a Streett automaton with $n$ states and $k$ pairs he constructs
a Rabin automaton with $(nk)^{O(nk)}$ states and $O(nk)$ pairs.

% Comment 26 - Orna
We mentioned that the Rabin and Streett conditions are duals;
the dual of the parity condition is parity again.
%Another issue with these stronger acceptance conditions is that the
%negation of one acceptance condition is not necessarily of the same
%type.
%In particular, the complement of Rabin is Streett but the complement
%of parity is parity.
Sometimes, given a nondeterministic automaton, we need to generate 
a deterministic automaton for the complementary language, a process
called {\em co-determinization} (e.g., for converting alternating tree
automata to nondeterministic tree automata). 
While complementing a deterministic automaton can be easily done by
dualizing the acceptance condition, such a dualization for a Rabin or
Streett automaton results in an automaton of the second type. 
%We call the process of constructing a deterministic automaton for the
%complement language co-determinization.
Thus, co-determinization of a B\"uchi (or Streett) automaton 
results in a deterministic Streett automaton.
Translating from Streett to Rabin or parity is exponential, 
we add a gadget with $k^2k!$ states where $k$ is the number
of pairs of the Streett condition \cite{Saf92}.
Thus, we multiply the number of states of the automaton by $k^2k!$.
\shorten{
The translation of Rabin to Streett or parity is dual and
has exactly the same complexity.
}{}

% Comment 28 - orna
Determinization has many uses other than complementation.
For example, Rabin uses McNaughton's determinization of B\"uchi automata to
complement nondeterministic Rabin tree automata
\cite{Rab72}.\footnote{% 
Rabin uses this complementation in order to prove that satisfiability
of S2S is decidable \cite{Rab72}.
This is essentially the same use that B\"uchi had for the
complementation of B\"uchi automata. In the context of tree
automata one has to use a more general acceptance condition.%
}
A node in an infinite tree belongs to infinitely many branches.
A tree automaton has to choose states that handle all branches in a
single run.
In many cases, we want all branches of the tree to belong to some word
language.
If we have a deterministic automaton for this word language, we run 
it in all directions simultaneously.
This kind of reasoning enables 
conversion of alternating tree automata to nondeterministic tree
automata and complementation of nondeterministic tree automata
(cf. \cite{Rab72,Tho90,Var98}).

Deterministic automata are used also for solving games and
synthesizing strategies.
In the context of games, the opponent may be able to choose between
different options. 
Using a deterministic automaton we can follow the game step by step
and monitor the goal of the game.
For example, in order to solve a game in which the goal is an LTL
formula, one first converts the LTL formula to a deterministic
automaton and then solves the resulting Rabin game \cite{PR89a}
(cf. \cite{KV98c,AHM01}).
Using Safra's determinization, reasoning about tree automata
reduces to reasoning about nondeterministic Rabin tree automata and
reasoning about general games reduces to reasoning about Rabin games.
Some of these applications use co-determinization, the deterministic
automaton for the complementary language.

In this paper we revisit Safra's determinization constructions. 
% In order to understand this you need to introduce Safra - orna
We show that we can further compact the tree structure used by Safra
to get a smaller representation of the deterministic automata.
By using dynamic node names instead of the static names used by Safra
we can construct directly a deterministic parity automaton.
Specifically, starting from a nondeterministic B\"uchi automaton
with $n$ states, we end up with a deterministic parity automaton with
$2n^nn!$ states and index $2n$ (instead of Rabin automaton with
$\safrastates$ states and $n$ pairs).
Starting from a Streett automaton with $n$ states and index $k$,
we end up with a deterministic parity automaton with
$\dpwsstates$ states and index $2n(k+1)$ (instead of Rabin automaton
with $\safrasstates$ states and $n(k+1)$ pairs). 
For both constructions, complementation is done by considering the
same automaton with a dual parity condition.

% comment 31 - orna
Though dividing the number of states by $12^n$ is not negligible, the
main importance of our result is in the fact that the resulting 
automaton is a parity automaton instead of Rabin.
Solving Rabin games (equivalently, emptiness of nondeterministic Rabin
tree automata) is NP-complete in the number of pairs \cite{EJ88}.
Solution of parity games is in NP$\cap$co-NP.
The current best upper bound for solving Rabin games is $mn^{k+1}k!$
where $m$ is the number of transitions, $n$ the number of states, and
$k$ the number of pairs \cite{PP06}.
Using our determinization construction instead of reasoning about
Rabin conditions we can consider parity conditions.
The best upper bound for solving parity games is $mn^{k/2}$
\cite{Jur00} (cf. \cite{BSV03,JPZ06} for other solutions).
\shorten{That is, we}{We} save a multiplier of at least $kk!$.

The gain by using our determinization is even greater when we consider
applications that use co-determinization.
As Streett is the dual of Rabin it follows that solving Streett
games is co-NP-complete.
Even if we ignore the computational difficulty, the Rabin acceptance
condition at least allows using memoryless strategies.
That is, when reasoning about Rabin games (or Rabin tree automata) the
way to resolve nondeterminism relies solely on the current location.
This is not the case for Streett.
In order to solve Streett games we require exponential memory
\cite{DJW97,Hor05}. 
% specify applications - orna
Applications like nondeterminization of alternating tree automata
use co-determinization but require the result to be a Rabin or parity 
automaton. 
Hence, the resulting deterministic Streett automaton has to be
converted to a parity automaton.
%Hence, the resulting deterministic Streett automaton is not
%good enough and it has to be converted to a parity automaton.
Again, the price tag of this conversion is a blowup of $k^2k!$ where
$k$ is the number of pairs.
% Not clear / badly written - orna
As the complexity of reasoning about parity games is $mn^{k/2}$, 
the extra multiplier grows to $(k^2k!)^k$. 
%%%Thus, in applications involving co-determinization the price tag of
%%%the Rabin/Streett acceptance condition is actually $(k^2k!)^k$.
%In the case that we start with a B\"uchi automaton with $n$ states
%this translates to saving a blowup of $(nn!)^n$.
%In the case that we start with a Streett automaton with $n$ states and
%$k$ pairs this translates to saving a blowup of
%$(n(k+1))^{n(k+1)} (n(k+1)!)^{n(k+1)}$.
%%Obviously, using our construction we do not have to pay this
%%additional price.

Recently, Kupferman and Vardi showed that they can check the emptiness
of an alternating parity tree automaton without directly using Safra's
determinization \cite{KV05c}.
Their construction can be used for many game / tree automata
applications that require determinization.
However, Kupferman and Vardi use Safra's determinization to get a
bound on the size of the minimal model of the alternating tree
automaton. 
Given such a bound, they can check emptiness by restricting the
search to small models.
Our improved construction implies that the complexity of their
algorithm reduces from  
$(12)^{n^2}n^{4n^2+2n}(n!)^{2n}$ to $(2n^nn!)^{2n}$.

\section{Nondeterministic Automata}
\Secl{defs}

\shorten{
Given a finite set $\Sigma$, a {\em word} over $\Sigma$ is a finite or
infinite sequence of symbols from $\Sigma$. 
We denote by $\Sigma^*$ the set of finite sequences over $\Sigma$ and
by $\Sigma^\omega$ the set of infinite sequences over $\Sigma$. 
Given a word $w = \sigma_0\sigma_1\sigma_2\cdots \in \Sigma^* \cup
\Sigma^\omega$, we denote by $w[i,j]$ the word $\sigma_i\cdots
\sigma_j$.
}{}

A {\em nondeterministic automaton} is $N= \zug{\Sigma, S, \delta, s_0,
\alpha}$, where $\Sigma$ is a finite alphabet, $S$ is a finite set of
states, $\delta: S \times \Sigma \rightarrow 2^S$ is a transition
function, $s_0 \in S$ is an initial state, and $\alpha$ is an acceptance
condition to be defined below.
A {\em run} of $N$ on a word $w=w_0w_1\cdots$ is an infinite sequence
of states $s_0 s_1 \cdots \in S^\omega$ such that $s_0$ is the initial
state and for all $j\geq 0$ we have $s_{j+1} \in \delta(s_j,w_j)$.
For a run $r=s_0s_1\cdots$, let $\inf(r) = \{s \in S ~|~ s=s_i \mbox{
  for infinitely many } i \mbox{'s}\}$ be the set of all states
occurring infinitely often in the run.
We consider four acceptance conditions. 
A {\em Rabin} condition $\alpha$ is a set of pairs $\{\zug{E_1,F_1},
\ldots, \zug{E_k,F_k}\}$ where for all $i$ we have \shorten{$E_i\subseteq S$ and
$F_i \subseteq S$}{$E_i,F_i\subseteq S$}. 
We call $k$ the {\em index} of the Rabin condition. 
A run is {\em accepting} according to the Rabin condition $\alpha$ if there
exists some $i$ such that $\inf(r) \cap E_i = \emptyset$ and $\inf(r)
\cap F_i \neq \emptyset$. 
That is, the run visits finitely often states from $E_i$ and
infinitely often states from $F_i$. 
The {\em Streett} condition is the dual of the Rabin condition.
Formally, a {\em Streett} condition $\alpha$ is also a set of pairs
$\{\zug{R_1,G_1},\ldots, \zug{R_k,G_k}\}$ where for all $i$ we have
\shorten{$R_i \subseteq S$ and $G_i \subseteq S$}{$R_i,G_i \subseteq
  S$}. 
We call $k$ the {\em index} of the Streett condition.
A run is {\em accepting} according to the Streett condition $\alpha$ if for
every $i$ either $\inf(r) \cap G_i = \emptyset$ or $\inf(r) \cap R_i
\neq \emptyset$. 
That is, the run either visits $G_i$ finitely often or visits $R_i$
infinitely often.
As a convention for pairs in a Rabin condition we use $E$ and $F$ and
for pairs in a Streett condition we use $R$ and $G$.
A {\em parity} condition $\alpha$ is a partition $\{F_0,\ldots, F_k\}$ of
$S$. 
We call $k$ the {\em index} of the parity condition.
A run is {\em accepting} according to the parity condition $\alpha$ if
for some even $i$ we have $\inf(r)\cap F_i \neq \emptyset$ and for all
$i'<i$ we have $\inf(r)\cap F_{i'}=\emptyset$.
A {\em B\"uchi} condition $\alpha$ is a subset of $S$. 
A run is {\em accepting} according to the B\"uchi condition $\alpha$ if
$\inf(r) \cap \alpha \neq \emptyset$. 
That is, the run visits infinitely often states from $\alpha$.
A word $w$ is {\em accepted} by $N$ if there exists some accepting run of
$N$ over $w$. 
The {\em language} of $N$ is the set of words accepted by
$N$. 
Formally, $L(N)=\{w ~|~ w \mbox{ is accepted by } N\}$. 
Two automata are {\em equivalent} if they accept the same language.

Given a set of states $S' \subseteq S$ and a letter $\sigma \in
\Sigma$, we denote by $\delta(S',\sigma)$ the set $\bigcup_{s\in
  S'}\delta(s,\sigma)$. 
Similarly, for a word $w \in \Sigma^*$ we define $\delta(S',w)$ in the
natural way: $\delta(S',\epsilon) = S'$ and $\delta(S',w\sigma) =
\delta(\delta(S',w),\sigma)$. 
For two states $s$ and $t$ and $w\in \Sigma^*$, we say that $t$ is
{\em reachable from $s$ reading $w$} if $t\in \delta(\{s\},w)$. 

An automaton is {\em deterministic} if for every state $s\in S$ and
letter $\sigma \in \Sigma$ we have $|\delta(s,\sigma)|=1$. 
In that case we write $\delta: S \times \Sigma \rightarrow S$.
\shorten{
We use deterministic automata to {\em complement} a word automaton,
i.e., construct an automaton that accepts the complement
language. 
We can use deterministic automata also as {\em monitors} in
games (see below). 
{\em Determinization} for automata on finite words is relatively
simple \cite{RS59}.
For automata on infinite words this is not the case.
Deterministic B\"uchi automata are strictly weaker than
nondeterministic B\"uchi automata. 
However, for every nondeterministic B\"uchi automaton there exists an
equivalent deterministic automaton with one of the stronger acceptance
conditions.
The best determinization constructions take nondeterministic B\"uchi
or Streett automata and convert them to deterministic Rabin
automata.
We describe these two constructions below.}{}

We use acronyms in $\{N,D\} {\times} \{R,S,P,B\} {\times}
\{W\}$ to denote automata.
The first symbol stands for the branching mode of the automaton: $N$
for nondeterministic and $D$ for deterministic. 
The second symbol stands for the acceptance condition of the
automaton: $R$ for Rabin, $S$ for Streett, $P$ for parity, and $B$ for
B\"uchi. 
The last symbol stands for the object the automaton is reading, in our
case $W$ for words.
For example, a DRW is a deterministic Rabin word automaton and an
NBW is a nondeterministic B\"uchi word automaton.

\section{Determinization of B\"uchi Automata}
\Secl{dpw}

In this section we give a short exposition of Safra's determinization
\cite{Saf88} and show how to improve it.
We replace the constant node names with dynamic names, which allow us
to simulate the {\em index appearance record}\footnote{%
The index appearance record is the gadget that allows to translate
Rabin and Streett conditions to parity conditions \cite{Saf92}.
It is a permutation over the pairs in the Rabin / Streett condition
with two pointers into the permutation.%
}
construction within the deterministic automaton.
We get a deterministic automaton with fewer states and in addition a
parity automaton instead of Rabin.

\subsection{Safra's Construction}

\shorten{
Here we describe Safra's determinization construction
\cite{Saf88,Saf89}. 
The construction takes an NBW and constructs an equivalent DRW. 
%This section is largely copied from \cite{Jut97,Lod98}.
}{
Here we describe Safra's determinization, which takes an NBW and
constructs an equivalent DRW \cite{Saf88}.}
Safra constructs a tree of subset constructions.
Every node in the tree is labeled by the states it follows.
The labels of siblings are disjoint and the label of a node is a
strict superset of the union of the labels of its descendants.
The sons are ordered according to their age.
The transition of a tree replaces the label of every node by the set
of possible successors. 
If the label now includes some accepting states, we add a new son to
the node with all these accepting states.
Intuitively, the states that label the sons of a node have already
visited an accepting state. Thus, the states in the label of a node
that are not in the labels of its descendants are states that still
owe a visit to the acceptance set.
If a state occurs in two sibling nodes (or more), we remove it from
the younger sibling and keep it only in the older sibling.
If the label of a node becomes equal to the union of labels of its
descendants then we mark this node as accepting and remove all its
descendants.
If some node remains eventually always in the tree and is marked
accepting infinitely often, the run is accepting.
Formally, we have the following.

Let $\cN = \zug{\Sigma, S, \delta, s_0, \alpha}$ be an NBW with
$|S| = n$. 
Let $V=[n]$. 
We first define Safra trees.

\begin{defi}
A {\em Safra tree} $t$ over $S$ is $\zug{N, r, p, \psi, l, E, F}$
where the components of $t$ are as follows.
\begin{enumerate}[$\bullet$]
\item
  $N\subseteq V$ is a set of nodes.
\item
  $r \in N$ is the root node.
\item
  $p: N \rightarrow N$ is the parent function defined over $N
  {-}\{r\}$, defining for every $v \in N {-} \{r\}$ its
  parent $p(v)$. 
  \shorten{}{We call children of the same node {\em siblings}.} 
\item
  $\psi$ is a partial order defining ``older than'' on 
  \shorten{siblings (i.e., children of the same node).}{siblings.}
\item
  $l : N \rightarrow 2^S$ is a labeling of the nodes with subsets of
  $S$. 
  The label of every node is a proper superset of the union of the
  labels of its sons. 
  The labels of two siblings are disjoint.
\item
  \shorten{
  $E,F \subseteq V$ are two disjoint subsets of $V$. They are used to
  define the Rabin acceptance condition.
  }
  {
  $E,F \subseteq V$ are disjoint and are used to define the Rabin
  condition.
  }
\end{enumerate}
\end{defi}

\noindent
The following claim is proved in \cite{Saf88,Saf89,Jut97,KV05c}.
\begin{claim}\Clml{Safra trees}
%  The number of nodes in a Safra tree is at most $n$. 
The number of Safra trees over $S$ is not more than $\safrastates$.
\end{claim}
\newcommand{\proofone}{
\myproof{1}{%
%  As the labels of siblings are disjoint and the union of labels of
%  children is a proper subset of the label of the parent it follows that
%  every node is the minimal (according to the subset order on the
%  labels) to contain (at least) some state $s \in S$. 
%  It follows that $n$ node names are sufficient.
%  
  The number of ordered trees on $n$ nodes is the $(n-1)$th Catalan number. 
  We know that $Cat(n)=\frac{(2n)!}{n!(n+1)!}$ and $Cat(n-1) \leq 4^n$.
  We represent the naming of nodes by $f:[n]\rightarrow [n]$ that
  associates the $i$th node with its name $f(i)$.
  There are at most $n^n$ such functions. 
  The labeling function is $l:S \rightarrow [n]$ where $l(s)=i$ means
  that $s$ belongs to the $i$th node and all its ancestors.
  Finally, we represent $E$ and $F$ by a function $a:V \rightarrow
  \{\emptyset,E,F\}$ such that $a(i)=\emptyset$ means that $i \notin E
  \cup F$, $a(i)=E$ means that $i \in E$, and $a(i)=F$ means that $i \in F$.
  There are at most $3^n$ such functions.
  
  To summarize, the number of trees is at most 
  $4^{n} \cdot 3^{n} \cdot n^n \cdot n^n = \safrastates$.
} % \newcommand{\proofone}
}
\proofone

We construct the DRW $\cD$ equivalent to $\cN$. 
Let $\cD = \zug{\Sigma, D, \rho, d_0, \alpha'}$ where the
components of $\cD$ are as follows.
\begin{enumerate}[$\bullet$]
\item 
  $D$ is the set of Safra trees over $S$. 
  For a state $d\in D$ we denote by a $d$ subscript the components of
  $d$. For example, $N_d$ is the set of nodes of $d$ and $l_d$ is the
  labeling of $d$.
\item
  \shorten{
  $d_0$ is the tree with a single node $1$ labeled by $\{s_0\}$ where
  $E$ is $V-\{1\}$ and $F$ is the empty set.
  }{
    $d_0$ is the tree with a single node $1$ labeled $\{s_0\}$ where
    $E{=}V{-}\{1\}$ and $F{=}\emptyset$.
  }
\item
  Let $\alpha'=\{\zug{E_1,F_1},\ldots, \zug{E_n,F_n}\}$ be the Rabin
  acceptance condition where
  $E_i=\{d\in D~|~i\in E_d\}$ 
  and  
  $F_i=\{d\in D~|~i\in F_d\}$. 
\item
  For every tree $d\in D$ and letter $\sigma \in \Sigma$ the
  transition $d'=\rho(d,\sigma)$ is the result of the following
  transformations on $d$. We use temporarily the set of names $V'$
  disjoint from $V$.
  
  \begin{enumerate}[(1)]
  \item
    \shorten{
    For every node $v$ with label $S'$ replace $S'$ by
    $\delta(S',\sigma)$ and set $E$ and $F$ to the empty set.
    }{
      For every node $v$ with label $S'$ replace $S'$ by
      $\delta(S',\sigma)$. Let $E=F=\emptyset$.
    }
  \item
    \shorten{
    For every node $v$ with label $S'$ such that $S' \cap \alpha \neq
    \emptyset$, create a new node $v' \in V'$ which becomes the
    youngest child of $v$. 
    Set its label to be $S' \cap \alpha$.
    }{
      For every node $v$ with label $S'$ such that $S' \cap \alpha \neq
    \emptyset$, create a new youngest child $v' \in V'$ of $v$.
    Set its label to $S' \cap \alpha$.
    }
  \item
    For every node $v$ with label $S'$ and state $s\in S'$ such that
    $s$ also belongs to the label of an older sibling $v'$ of $v$,
    remove $s$ from the label of $v$ and all its descendants.
    \shorten{
  \item
    Remove all nodes with empty labels.}{}
  \item
    For every node $v$ whose label is equal to the union of the labels
    of its children, remove all descendants of $v$. 
    Add $v$ to $F$.
  \item
    \shorten{
    Add all unused nodes to $E$.}
    {Remove all nodes with empty labels and add all unused names to $E$.}
  \item
    Change the nodes in $V'$ to nodes in $V$.\label{changing the names}
  \end{enumerate}
\end{enumerate}

\begin{claim}
The transition is well defined.
\end{claim}

\begin{proof}
It is simple to see that the label of every node is a proper superset
of the union of the labels of its children and that the labels of two
siblings are disjoint.
We have to show that the $n$ nodes in $V$ are sufficient to complete
step 7.%\ref{changing the names}.
As the labels of siblings are disjoint and the union of labels of
children is a proper subset of the label of the parent it follows that
every node is the minimal (according to the subset order on the
labels) to contain (at least) some state $s \in S$. 
It follows that $n$ node names are sufficient.
\end{proof}

\begin{thm}{\rm \cite{Saf88}}
$L(\cD) = L(\cN)$. \qed
\end{thm}

\noindent
For other expositions of this determinization \shorten{we refer the
  reader to}{confer}
\cite{Jut97,Lod98,Rog01}.

\subsection{From NBW to DPW}

We now present our construction.
Intuitively, we take Safra's construction and replace the constant
node name with a dynamic one that decreases as nodes below it get
erased from the tree.
Using the new names we can give up the ``older than'' relation.
The smaller the name of a node, the older it is.
Furthermore, the names give a natural parity order on
good and bad events. 
Erasing a node is a bad event (which forces all nodes with greater
name to change their name).
Finding that the label of some name is equal to the union of labels of
its descendants is a good event.
The key observation is that a node can change its name 
\shorten{at most a}{only a}
finite number of times without being erased.
\shorten{It follows that the}{The} names of all nodes that stay
eventually in the tree get constant. 
Thus, bad events happen eventually only to nodes that get erased from
the tree.
Then we can monitor good events that happen to the nodes with constant
names and insist that they happen infinitely often. 
Formally, we have the following.

Let $\cN = \zug{\Sigma, S, \delta, s_0, \alpha}$ be an NBW with $|S|=n$.
For the sake of the proof we would like to treat the nodes as 
entities. 
Hence, we distinguish between the set of nodes $V=[2n]$ of a tree and
their names that may change and range over $[n]$.
All important information (tree structure, label) can
be associated with the names and in practice the distinction between
nodes and their names is not needed.

\begin{defi}
A {\em compact Safra tree} $t$ over $S$ is $\zug{N, M, 1, p, l, e,
  f}$ where the components of $t$ are as follows. 
\begin{enumerate}[$\bullet$]
\item
  $N \subseteq V$ is a set of nodes.
\item
  $M: N \rightarrow [n]$ is the naming function.
\item
  $1\in N$ such that $M(1)=1$ is the root node.
\item
  $p:N \rightarrow N$ is the parent function.
\item
  $l:N \rightarrow 2^S$ is a labeling of the nodes with subsets of
  $S$.
  The label of every node is a proper superset of the union of the
  labels of its sons. 
  The labels of two siblings are disjoint.
\item
  $e,f \in [n+1]$ are used to define the parity acceptance condition.
  The number $e$ is used to memorize the minimal node that changed
  its name and $f$ the minimal node that is equivalent to its
  descendants.
\end{enumerate}
\end{defi}

\noindent
Notice that we give up the ``older than'' relation and replace
the sets $E$ and $F$ by numbers $e$ and $f$.
We require that the naming $M$ is a bijection from $N$ to $[|N|]$. 
That is, the names of the nodes in $N$ are consecutive
starting from the root, which is named $1$.

\shorten{
The following claim is proved much like the similar proof for Safra
trees.}{}
\begin{claim}\Clml{compact safra trees}
  The number of compact Safra trees over $S$ is not more than
  $2n^nn!$. 
\end{claim}
\newcommand{\proofthree}{
\myproof{3}{
  Just like Safra trees there are at most $n$ nodes.
  We use only the names of the nodes. 
  The parent of the node has a smaller name. 
  Thus, the parenthood relation can be represented by a sequence of
  (at most) $n-1$ pointers, where the $i$th pointer is pointing to a
  value in $1,\ldots, i{-}1$.
  It follows that there are at most $(n{-}1)!$ such trees.
  As in Safra trees, every node has at least one unique state in $S$
  that belongs to it.  
  We add the function $l:S \rightarrow [n]$ that associates a state
  with the minimal node (according to the descendant order in the
  tree) to which it belongs.
  There are $n$ options for $e$ and $f$ each.
  In order to define the acceptance condition (see below) we need to
  know the value of $e$ in case $e\leq f$ and the value of $f$ in case
  $f<e$.
  Thus, we need $2n$ possible values.
  It follows that there are at most 
  $2n \cdot (n{-}1)! \cdot n^n=2n^nn!$
  different compact Safra trees.
} % end of proof 3
} %\newcommand{\proofthree}
\proofthree

\noindent
We construct the DPW $\cD$ equivalent to $\cN$.
Let $\cD = \zug{\Sigma, D,\rho, d_0, \alpha'}$ where the components of
$\cD$ are as follows.

\begin{enumerate}[$\bullet$]
\item
  $D$ is the set of compact Safra trees over $S$.
\item
  \shorten{
    $d_0$ is the tree with a single node $1$ labeled $\{s_0\}$ and named
    $1$ where $e=2$ and $f=1$.
  }{
    $d_0 \in D$ has a unique node $1$ where $l(1){=}\{s_0\}$,
    $M(1){=}1$, $e{=}2$, and $f{=}1$.
  }
\item
  The parity acceptance condition $\alpha'{=}\zug{F_0,\ldots, F_{2n-1}}$
  is defined as follows.
  \begin{enumerate}[-]
  \item
    $F_0 = \{ d \in D ~|~ f=1 \mbox { and } e > 1 \}$
  \item
    $F_{2i+1} = \{ d\in D ~|~ e=i+2 \mbox{ and } f \geq e\}$
  \item
    $F_{2i+2} = \{ d\in D ~|~ f=i+2 \mbox{ and } e > f \}$
  \end{enumerate}
  Note that the case $e=1$ is not considered above. 
  In this case the label of the root is empty.
  This is a rejecting sink state.
\item
  For every tree $d\in D$ and letter $\sigma\in \Sigma$ the transition
  $d'=\rho(d,\sigma)$ is the result of the following transformations
  on $d$.

  \begin{enumerate}[(1)]
  \item
    For every node $v$ with label $S'$ replace $S'$ by
    $\delta(S',\sigma)$.
  \item
    For every node $v$ with label $S'$ such that $S' \cap \alpha \neq
    \emptyset$, create a new son $v' \notin N$ of $v$.
    Set its label to $S' \cap \alpha$. 
    Set its name to the minimal value greater than all used names.
    We may have to use temporarily names in the range $[(n{+}1)..(2n)]$.
  \item
    For every node $v$ with label $S'$ and state $s\in S'$ such that
    $s$ belongs also to some sibling $v'$ of $v$ such that
    $M(v')<M(v)$, remove $s$ from the label of $v$ and all its
    descendants.
  \item
    \label{step in transition}
    For every node $v$ whose label is equal to the union of the labels
    of its children, remove all descendants of $v$. 
    Call such nodes {\em green}. 
    Set $f$ to the minimum of $n{+}1$ and the names of green nodes.
    Notice that no node in $[(n{+}1)..(2n)]$ can be green.
  \item
    Remove all nodes with empty labels. 
    Set $e$ to the minimum of $n{+}1$ and the names of nodes removed
    during all stages of the transformation.
    Notice that the priority of a state is even only when $f < e$.
    Thus, green nodes that are removed cannot make a state of even
    priority.
  \item
    Let $Z$ denote the set of nodes removed during all previous
    stages of the transformation. 
    For every node $v$ let $rem(v)$ be $|\{v' \in Z ~|~
    M(v')<M(v)\}|$. 
    That is, we count how many nodes are removed during the
    transformation and have smaller name than the name of $v$.
    For every node $v$ such that $l(v)\neq \emptyset$ we change the
    name of $v$ to $M(v)-rem(v)$.
    It is simple to see that the resulting names are consecutive
    again and in the range $[n]$.%\footnote{
%    Suppose that two nodes are named $p'>p$ before the
%    name change and $p''$ after the name change.
%    This implies that $p-p''$ nodes with name smaller than $p$ are
%    removed and $p'-p''$ nodes with name smaller than $p'$ are
%    removed. Thus, the number of nodes removed whose name is between
%    $p$ and $p'$ is $p'-p$, which implies that the node named $p$
%    itself is removed.}
  \end{enumerate}
\end{enumerate}

\shorten{
\noindent
We show that the two automata are equivalent.
The proof is an adaptation of Safra's proof \cite{Saf88}.

\begin{thm}
  $L(\cD) = L(\cN)$.
  \Thml{language equivalence DPW}
\end{thm}
}
{
%The proof that the two automata are equivalent resembles Safra's proof
%\cite{Saf88} and is given in ???.
}

\newcommand{\prooffour}{
\myproof{4}{
  Consider $w\in L(\cN)$. 
  We have to show $w\in L(\cD)$. 
  Let $r = s_0s_1\cdots$ be an accepting run of $\cN$ on $w$.
  Let $r' = d_0d_1\cdots$ be the run of $\cD$ on $w$ and 
  let $d_i = \zug{N_i,M_i,1,p_i,l_i,e_i,f_i}$.
  It is simple to see that for all $i\geq 0$ we have $s_i \in l_i(1)$
  and $e_i>1$.
%  If step~\ref{step in transition} is applied infinitely often to node
  If step 4 is applied infinitely often to node
  1 (equivalently, $f=1$ infinitely often, or during the
  transformation of the trees the label of $1$ equals the labels of its
  sons) then $r'$ visits $F_0$ infinitely often.

%  Otherwise, from some point onwards in $r'$ we have step~\ref{step in
%  transition} is not applied to node $1$.
  Otherwise, from some point onwards in $r'$ we have step 4
  is not applied to node $1$.
  Let $i_0$ be this point.
  There exists a point $i_1>i_0$ such that $s_{i_1} \in \alpha$. 
  It follows that for all $i>i_1$ we have $s_{i}$ belongs to some
  son $v_1$ of $1$.
  Notice, that just like in Safra's case, the run $r$ may start in
  some son of $1$ and move to a son with a smaller name. 
  However, this can happen finitely often and hence we treat $v_1$ as
  constant. 
  The name $M(v_1)$ may decrease finitely often until it is constant. 
  Let $i_2$ be such that for all $i>i_2$ we have $a_1=M_{i}(v_1)$. 
  As $M_{i}(v_1)=a_1$ for all $i>i_2$ it follows that $e_{i}>a_1$
  for all $i>i_2$.

%  Suppose that step~\ref{step in transition} is applied to $v_1$
  Suppose that step 4 is applied to $v_1$
  infinitely often (equivalently, $f\leq a_1$ infinitely often).
  It follows that for every odd $a'<2a_1-2$ we have $F_{a'}$ is visited
  finitely often and either $F_{2a_1-2}$ is visited infinitely often
  or there exists some even $a'<2a_1-2$  such that $F_{a'}$ is visited
  infinitely often. In this case $\cD$ accepts $w$.

%  Otherwise, step~\ref{step in transition} is applied to $v_1$
  Otherwise, step 4 is applied to $v_1$
  finitely often. 
  We construct by induction a sequence $v_1, \ldots, v_k$ such
  that eventually $v_1,\ldots, v_k$ do not change their names and $r$
  belongs to all of them.
  As the number of active nodes in a tree (nodes $v$ such that
  $l(v)\neq \emptyset$) is bounded by $n$ we can repeat the process
  only finitely often.
  Hence, $w$ is accepted by $\cD$.
  
  In the other direction, consider $w\in L(\cD)$. 
  Let $r' = d_0d_1\cdots$ be the accepting run of $\cD$ on $w$ where
  $d_i=\zug{N_i,M_i,1,p_i,l_i,f_i,e_i}$.
  Let $F_{2a}$ be the minimal set to be visited infinitely often.
  It follows that eventually always $e_i > a+1$ and infinitely often
  $f_i=a+1$.
  We first prove two claims. 
%  The first, showing that all the states of $\cN$ that appear in
%  labels of nodes of a state of $\cD$ are reachable from the initial
%  state of $\cN$. 
%  The second, proves that if for some $2j$ the set $F_{2j}$ is
%  visited in $d_i$ and in $d_{i'}$ and no visit to $F_{j'}$ for $j'<2j$
%  occurs between $i$ and $i'$, then there exists a node $v$ such that
%  $M_a(v)=j+1$ for all $i \leq a \leq i'$ and for every state $s$ in
%  $l_{i'}(v)$ we  find a run segment of $\cN$ that starts from some
%  state of $l_i(v)$, visits $\alpha$, and ends in $s$.

  \begin{claim}
    For every $i \in \mathbb{N}$, $v\in N_i$, and every state $s\in
    l_i(v)$ we have $s$ is reachable from $s_0$ reading $w[0,i-1]$.
    \Clml{reachability dpw}
  \end{claim}

  \myproof{4.1}{ 
    We prove the claim simultaneously for all $v \in N_i$ by induction
    on $i$. 
    Clearly, it holds for $i=0$.
    Suppose that it holds for $i$.
    As $l_{i+1}(v) \subseteq \delta(l_i(v'),w_i)$ for some $v' \in N_i$  
    it follows that every state in $l_{i+1}(v)$ is reachable from $s_0$
    reading $w[0,i]$.
  } % end of proof 4.1

  \begin{claim}
    Consider $i,i' \in \mathbb{N}$ such that $i<i'$,
    $d_i,d_{i'} \in F_{2a}$ for some $a$, and for all $a' \leq 2a$ and
    for all $i < j < i'$ we have $d_{j} \notin F_{a'}$.
    Then there exists a node $v$ such that $M_{j}(v)=a+1$ for all
    $i\leq j \leq i'$ and every state $s$ in $l_{i'}(v)$ is reachable
    from some state in $l_i(v)$ reading $w[i,i'-1]$ with a run that
    visits $\alpha$.
    \Clml{visit to F dpw}
  \end{claim}

  \myproof{4.2}{
    There exists some node $v$ such that $M_i(v)=a+1$ (as $d_i \in
    F_{2a}$).
    By assumption, for every $a'<2a$ the set $F_{a'}$ is not visited
    between $i$ and $i'$.    
    Hence, for every node $v'$ such that $M_i(v)\leq a+1$ we have that
    $M_j(v')=M_i(v')$ for all $i\leq j \leq i'$.
    That is, between $i$ and $i'$ all nodes whose name is at most
    $a+1$ do not change their names.
    In particular, for all $i\leq j \leq i'$ we have $M_j(v)=a+1$.
%    If $i'=i+1$ then all the states in $l_{i'}(v)$ are in $\alpha$ and we
%    are done. 
%    Otherwise, $i' > i+1$.
    We show that for every $i \leq j < i'$ and every
    descendant $v'$ of $v$, every state in $l_j(v')$
    is reachable from some state in
    $l_i(v)$ along a run visiting $\alpha$. 
    As $v$ is a leaf in $d_i$ for $j=i$ this is obviously true.
%    Consider some descendant $v'$ of $v$ appearing in $d_{i+1}$ (there
%    is at most one, it must exist as $F_{2j}$ is not visited between
%    $i$ and $i'$ and $i' > i+1$).
%    As $l_{i+1}(v') \subseteq \delta(l_i(v),w_i) \cap \alpha$ this is
%    obviously true for $i+1$.
    Suppose it is true for $j$ and prove for $j+1$.
    We know that for every descendant $v'$ of $v$ either $l_{j+1}(v')
    \subseteq \delta(l_{j}(v),w_j) \cap \alpha$ or for some descendant
    $v''$ of $v$ we have $l_{j+1}(v') \subseteq
    \delta(l_{j}(v''),w_j)$ ($v''$ may be $v'$). 
    As during the transformation from $d_{i'-1}$ to $d_{i'}$ the label
    $l_{i'}(v)$ equals the union of labels of sons of $v$ the claim
    follows.
    In particular, if $i'=i+1$ then all states in $l_{i'}(v)$ are in
    $\alpha$ and we are done.
  } % end of proof 4.2

  We construct an infinite tree with finite branching degree.
  The root of the tree corresponds to the initial state of $\cN$.
  Every node in the tree is labeled by some state of $\cN$ and a time
  stamp $i$.
  An edge between the nodes labeled $(s,i)$ and $(t,j)$ corresponds to
  a run starting in $s$, ending in $t$, reading $w[i,j-1]$, and
  visiting $\alpha$.
  From K\"onig's lemma this tree contains an infinite branch. 
  The composition of all the run segments in this infinite branch is
  an infinite accepting run of $\cN$ on $w$. 

%  We are now ready to build the tree $T$. 
  Let $(s_0,0)$ label the root of $T$. 
  Let $i_0$ be the maximal location such that for all $a'<2a$ the set
  $F_a$ is not visited after $i_0$.
  Let $v$ be the node such that for all $i>i_0$ we have $M_{i}(v)=a+1$.
  Let $i_1$ be the minimal location such that $i_1> i_0$ and
%  $f_{i_1}=a+1$ (that is step~\ref{step in transition} was applied to $v$).
  $f_{i_1}=a+1$ (that is step 4 was applied to $v$).
  For every state $s$ in $l_{i_1}(v)$ we add a node to $T$, label it
  by $(s,i_1)$ and connect it to the root. 
  We extend the tree by induction.
  We have a tree with leafs labeled by the states in $l_{i_j}(v)$
%  stamped by time $i_j$, and $f_{i_j}=a+1$ (step~\ref{step in transition}
  stamped by time $i_j$, and $f_{i_j}=a+1$ (step 4
  was applied to $v$).
  That is, for every state $s$ in $l_{i_j}(v)$ there exists a leaf
  labeled $(s,i_j)$.
  We know that $F_{2a}$ is visited infinitely often. 
  Hence, there exists $i_{j+1} > i_j$ such that $f_{i_{j+1}}=a+1$
%  (step~\ref{step in transition} is applied to $v$).
  (step 4 is applied to $v$).
  For every state $s'$ in $l_{i_{j+1}}(v)$ we add a node to $T$ and 
  label it $(s',i_{j+1})$. From \Clmr{visit to F dpw} there exists a state
  $s$ in $l_{i_j}(v)$ such that $s'$ is reachable from $s$ reading
  $w[i_j,i_{j+1}-1]$ with a run that visits $\alpha$. We connect
  $(s',i_{j+1})$ to $(s,i_j)$. 

  From \Clmr{reachability dpw} it follows that every edge
  $(s_0,0),(s',i_1)$ corresponds to some run starting in $s_0$, ending
  in $s'$, and reading $w[0,i_1-1]$. 
  From \Clmr{visit to F dpw}, every other edge in the tree
  $(s,i_j),(s',i_{j+1})$ corresponds to some run starting in $s$, ending in
  $s'$, reading $w[i_j,i_{j+1}-1]$, and visiting $\alpha$. 
  From K\"onig's lemma there exists an infinite branch in the tree.
  This infinite branch corresponds to an accepting run of $\cN$
  on $w$. 
} % end of proof 4
} % end of \newcommand{\prooffour}
\prooffour

\begin{thm}
  For every NBW $\cN$ with $n$ states there exists a DPW $\cD$ with
  $2n^nn!$ states and index $2n$ such that $L(\cD)=L(\cN)$. \qed
  \Thml{dpw complexity}
\end{thm}

\noindent
We note that this improves Safra's construction in two ways. 
First, we reduce the number of states from $\safrastates$ to
$2n^nn!$.
Second, our automaton is a parity automaton which is amenable to
simpler algorithms.
For example, given a DPW it is possible to check in polynomial time
what is the minimal parity index that enables recognition of the same
language, and to find such an optimal parity index on the same
automaton structure \cite{CM99}.
On the other hand, finding the minimal Rabin index of a DRW is NP-hard
\cite{KPBV95} and it may be the case that an optimal condition cannot
be found on the same structure \cite{KMM04}.
Many times we are interested in a deterministic automaton for the
complement language, a process called co-determinization.
The natural complement of a DRW is a DSW.
However,the Streett acceptance condition is less convenient in many
applications (due to the fact that Streett acceptance conditions
require memory).
Thus, the complement automaton is usually converted to a DPW using the
IAR construction \cite{Saf92}.
In such a case, one would have to multiply the number of states by
$k^2k!$ where $k$ is the number of Rabin pairs.
A similar effect occurs when using deterministic automata in the
context of games.
Solution of Rabin games incurs an additional multiplier of
$k^2k!$.
With our construction this penalty is avoided.

\section{Determinization of Streett Automata}
\Secl{dpw2}

\newcommand{\nswtodpwshortone}{
In this section we show that the ideas described in \Secr{dpw} can be
applied to Safra's determinization of Streett automata \cite{Saf92}.

}
\newcommand{\streettintuition}{
As mentioned, in the case of Streett automata, determinization via
conversion to B\"uchi automata is less than optimal.
Safra generalizes his construction to work for Streett automata.
The idea is still to use a set of subset constructions. 
Let $\cS = \zug{\Sigma, S, \delta, s_0, \alpha}$ be an NSW where $\alpha =
\{ \zug{R_1,G_1}, \ldots, \zug{R_k,G_k}\}$.
We say that a run $r$ of $\cS$ is accepting according to the {\em
witness set} $J \subseteq [k]$ if for every $j\in J$ we have $\inf(r)
\cap R_j \neq \emptyset$ and for every $j\notin J$ we have $\inf(r)
\cap G_j = \emptyset$.
It is easy to construct an NBW whose language consists of all words
accepted according to witness set $J$. 
The NBW has two parts. In the first part
it waits until all visits to $G_j$ for $j\notin J$ have occurred.
Then it moves nondeterministically to the second part where it waits
for visits to $R_j$ for each $j \in J$ according to their order and
disallows visits to $G_j$ for every $j\notin J$.
If the automaton loops through all $j\in J$ infinitely often the run
is accepting. 
Unfortunately, the number of possible witness sets is exponential.

Safra's construction arranges all possible runs of the NSW and all
relevant witness sets in a tree structure.
A state is again a tree of subset constructions. 
Every node in a tree represents a process that is monitoring some
witness set and checking this witness set.
The node for witness set $J$ follows some set of states. 
It waits for visits to $R_j$ for every $j\in J$ (in descending order),
if this happens without visiting $G_j$ for $j\notin J$ then the node
succeeds and starts all over again.

A {\em Streett Safra tree} is a tree whose nodes are labeled by
subsets of the states in $S$. The labels of siblings are disjoint and
the labels of sons form a partition of the label of the parent. 
In addition every node is annotated by a subset $J \subseteq [k]$.
The annotation of a son misses at most one element from the annotation
of the parent. Every node that is not a leaf has at least one son
whose annotation is a strict subset.
In addition, children are ordered according to their age.

The root node monitors the set $[k]$ as a possible witness set.
If some node is annotated with $J$ and has a child annotated $J-\{j\}$
this means that the child has given up on the hope that $R_j$ will
occur.
If a node has given up on $R_j$ but visits $G_j$ then the states
visiting $G_j$ have no place in this node and they are moved to a new
sibling.
Similarly, if a node has given up on $R_j$ and visits $R_j$ then the
states visiting $R_j$ have no place in this node and they are moved to
a new sibling.
Whenever the label of a node gets empty it is removed from the tree.
If all the states followed by a node completed a cycle through its
witness set, all the descendants of this node are removed and it is
marked accepting.
The Rabin condition associates a pair with every node.
A run is accepting if some node is erased finitely often and marked
accepting infinitely often.
} % end of \streettintuition
\newcommand\nswtodpwshorttwo{
For further details we refer the reader to \cite{Saf92,Jut97,Sch01}.

\begin{thm}{\rm \cite{Saf92}}
For every NSW $\cS$ with $n$ states and $k$ pairs there exists a DRW
$\cR$ with $\safrasstates$ and $n(k{+}1)$ pairs such that
$L(\cR)=L(\cS)$.
\end{thm}

As before, Safra uses constant node names to define the Rabin
condition. 
We replace this with dynamic node names that maintain an order between
all the nodes in the tree.
As before, the order between the nodes allows us to give an order to
good and bad events and we get a DPW.
}
\newcommand{\nswtodpwshort}{
\nswtodpwshortone
\streettintuition
\nswtodpwshorttwo
}

\newcommand{\nswtodpw}{
In this section we give a short exposition of Safra's determinization
of Streett automata \cite{Saf92} and show how to improve it.
Again, we replace the constant node names with dynamic names.
We get a deterministic automaton with fewer states and in addition a
parity automaton instead of Rabin.
The intuition is similar to the construction in \Secr{dpw}.

\subsection{Safra's Construction}

Here we describe Safra's determinization for Streett
Automata \cite{Saf92}.
The construction takes an NSW and constructs an equivalent DRW. 
%\shorten{This section is largely copied from \cite{Saf92,Jut97,Sch01}.}{}

\streettintuition

Let $\cS = \zug{\Sigma, S, \delta, s_0,\alpha}$ be an NSW where $\alpha =
\{\zug{R_1,G_1}, \ldots, \zug{R_k,G_k}\}$ and $|S|=n$.
Let $m=n(k+1)$ and $V = [m]$.
\shorten{
We first define Streett Safra trees.}{}

\begin{defi}
A {\em Streett Safra tree} $t$ over $S$ is
$\zug{N,r,p,\psi,l,h,E,F}$ where the components of $t$ are as follows.
\begin{enumerate}[$\bullet$]
\item
  $N \subseteq V$ is the set of nodes.
\item
  $r\in N$ is the root node.
\item
  $p:N \rightarrow N$ is the parent function defined over $N{-}
  \{r\}$, defining for every $v\in N{-} \{r\}$ its parent
  $p(v)$.
\item
  $\psi$ is a partial order defining ``older than'' on siblings (i.e.,
  children of the same node).
\item
  $l:N \rightarrow 2^S$ is a labeling of nodes with subsets of
  $S$. 
  The label of every node is equal to the union of the labels of
  its sons. 
  The labels of two siblings are disjoint.
\item
  $h:N \rightarrow 2^{[k]}$ annotates every node with a set of indices
  from $[k]$. 
  The root is annotated by $[k]$.
  The annotation of every node is contained in that of its parent and
  it misses at most one element from the annotation of the parent. 
  Every node that is not a leaf has at least one son with strictly
  smaller annotation. 
\item
  $E,F \subseteq V$ are two disjoint subsets of $V$. They are used to
  define the Rabin acceptance condition.
\end{enumerate}
\end{defi}

\noindent
The following claim is proved in \cite{Saf92,Sch01}.
\begin{claim}  \Clml{Streett Safra trees}%\hfill
The number of Streett Safra trees over $S$ is at most
\[\safrasstates.\] 
\end{claim}
\newcommand{\proofofstreettsafratrees}{
\myproof{2}{
%  The labeling is determined by the labels of the leaves.
%  As the labels of all leaves are disjoint there are at most $n$ leaves.
%  We can represent the annotation $h$ by annotating every edge by
%  the value $j \in [k]$ such that $j$ is in the annotation of the parent
%  and not in the annotation of the son. 
%  If no such $j$ exists then we annotate the edge by $0$.
%  There exists a path from the root to a leaf where no edge is
%  annotated by $0$. 
%  For every edge annotated 0, there is a path from the
%  target of this edge to a leaf where no edge is annotated by
%  $0$.
%  Hence, there are at most $n-1$ edges annotated by $0$. 
%  Every other edge is either annotated by some index $i \in [k]$ or
%  connects a parent $v$ to a son $v'$ such that there is some
%  state $s \in S$ such that $s\in l(v)$ and $s\notin l(v')$.
%  Thus, there can be at most $nk$ such edges.
%  Totally, $n(k+1)$ node names are sufficient.
  
  The number of ordered trees on $m$ nodes is at most $4^m$.
  We represent the naming of the nodes by $f:[m]\rightarrow [m]$.
  There are at most $m^m$ such functions.
  The labeling of a node is determined by the labels of the leaves in
  the subtree below it and labels of leaves are disjoint.
  The labeling function $S \rightarrow [n]$ associates a state
  $s$ with the leaf it belongs to.
  There are at most $n$ leaves and $n^n$ such functions.
  We can represent the annotation $h$ by annotating every edge by the
  value $j\in [k]$ such that $j$ is in the annotation of the parent
  and not in the annotation of the son.
  If no such $j$ exists then we annotate the edge by $0$.
  The edge annotation function is $h:[m] \rightarrow [0..k]$ associating
  an index to the target node of the edge. 
  Finally, $E$ and $F$ are represented by a function $a: V \rightarrow
  \{\emptyset,E,F\}$.
%
%  \noindent
  The number of trees is at most $4^{m} \cdot 3^{m} \cdot m^m \cdot
  n^n \cdot (k+1)^m = \safrasstates$.
} % end of proof 2 
}
\proofofstreettsafratrees

\noindent
We construct the DRW $\cD$ equivalent to $\cS$.
Let $\cD = \zug{\Sigma, D, \rho, d_0, \alpha'}$ where the components of
$\cD$ are as follows.
\begin{enumerate}[$\bullet$]
\item
  $D$ is the set of Streett Safra trees over $S$.
\item
  \shorten{
  $d_0$ is the tree with a single node $1$ labeled by $\{s_0\}$ where
  $E$ is $V-\{1\}$ and $F$ is the empty set.
  }{
    $d_0$ is the tree with a single node $1$ labeled $\{s_0\}$ where
  $E{=}V{-}\{1\}$ and $F{=}\emptyset$.
  }
\item
  Let $\alpha'=\{\zug{E_1,F_1},\ldots, \zug{E_m,F_m}\}$ be the Rabin
  acceptance condition where
%  The Rabin acceptance condition $\alpha'$ is 
%  $\{\zug{E_1,F_1}\ldots, \zug{E_{m},F_{m}}\}$ where 
%
  $E_i = \{d \in D ~|~ i \in E_d\}$ and $F_i = \{d \in D ~|~ i \in F_d\}$.
\item
  For every tree $d\in D$ and letter $\sigma \in \Sigma$ the
  transition $d'=\rho(d,\sigma)$ is the result of the following
  (recursive) transformation applied on $d$ starting from the root.
  Before we start, we set $E$ and $F$ to the empty set and replace the
  label of every node $v$ by $\delta(l(v),\sigma)$. We use temporarily
  the set of names $V'$ disjoint from $V$.
  
  \begin{enumerate}[(1)]
  \item
    If $v$ is a leaf such that $h(v)=\emptyset$ stop.
  \item
    If $v$ is a leaf such that $h(v) \neq \emptyset$, add to $v$ a new
    son $v' \in V'$.
    Set $l(v')=l(v)$ and $h(v') = h(v) - \{max(h(v))\}$.
    \label{item two}
  \item
    Let $v_1,\ldots, v_l$ be the sons of $v$ (ordered from oldest to
    youngest) and let $j_1, \ldots, j_l$ be the indices 
    such that $j_i \in h(v)-h(v_i)$ (note that $|h(v)-h(v_i)| \leq 1$;
    in case that $h(v)=h(v_i)$ we have $j_i=0$).
    Call the entire procedure recursively on $v_1,\ldots, v_l$
%    (call recursively also for sons created in step \ref{item two}
    (call recursively also for sons created in step 2
    above). 
    
    For every son $v_i$ and every state $s \in l(v_i)$ do the following.
    \begin{enumerate}[(a)]
    \item
      If $s\in R_{j_i}$, 
      remove $s$ from the label of $v_i$ and all its descendants.
      Add a new youngest son $v' \in V'$ to $v$.
      Set $l(v')=\{s\}$ and $h(v') = h(v)-\{max(\{0\} \cup (h(v) \cap
      \{1,\ldots,j_i-1\}))\}$.
    \item
      If $s \in G_{j_i}$,
      remove $s$ from the label of $v_i$ and all its descendants.
      Add a new youngest son $v'\in V'$ to $v$.
      Set $l(v')=\{s\}$ and $h(v') = h(v)-\{j_i\}$.%
      \footnote{%
        We note that in Safra's original construction
        \cite{Saf92,Sch01} the rank of the new node is set to
        $h(v')=h(v)-\{max(h(v))\}$. 
        In case that both $G_{j_i}$ and $R_{j_i}$ are visited
        infinitely often this may lead to the following situation. 
        Suppose that the node $v$ has a son $v'$ that is waiting for a
        visit to $R_{j_i}$ where $j_i$ is not the maximum in $h(v)$.  
        In the case that $G_{j_i}$ is visited, the runs are moved to
        new siblings that await $max(h(v))$ again.
        This way, the run may cycle infinitely often between
        $max(h(v))$ and $j_i$, leading to incompleteness of the
        construction.% 
    }
    \end{enumerate}
  \item
    If a state $s$ appears in $l(v_i)$ and $l(v_{i'})$ and $j_i
    < j_{i'}$ then remove $s$ from the label of $v_{i'}$ and all its
    descendants. 
  \item
    If a state $s$ appears in $l(v_i)$ and $l(v_{i'})$ and $j_i
    = j_{i'}$ then remove $s$ from the label of the younger sibling
    and all its descendants.
    \shorten{
  \item
    Remove sons with empty label.}{}
  \item
    If all sons \shorten{}{whose label is not empty} are annotated by
    $h(v)$ remove all the sons and all their descendants.
    Add $v$ to $F$.
  \end{enumerate}
  Finally, we \shorten{}{remove nodes with empty label,} add all
  unused names to $E$, remove unused names from $F$, and change the
  nodes in $V'$ to nodes in $V$.
\end{enumerate}

\begin{claim}
The transition is well defined.
\end{claim}

\begin{proof}
It is simple to see that the label of every node is equal to the union
of the labels of its children and that the labels of two siblings are
disjoint. 
We have to show that the $m$ nodes in $V$ are sufficient to change the
nodes in $V'$ to nodes in $V$.

There exists a path from the root to a leaf where no edge is
annotated by $0$. 
For every edge annotated 0, there is a path from the
target of this edge to a leaf where no edge is annotated by
$0$.
Hence, there are at most $n-1$ edges annotated by $0$. 
Every other edge is either annotated by some index $i \in [k]$ or
connects a parent $v$ to a son $v'$ such that there is some
state $s \in S$ such that $s\in l(v)$ and $s\notin l(v')$.
Thus, there can be at most $nk$ such edges.
Totally, $n(k+1)$ node names are sufficient.
\end{proof}

\begin{thm}{\rm \cite{Saf92}}
$L(\cD) = L(\cS)$.
\end{thm}

\noindent
For other expositions of this determinization \shorten{we refer the
  reader to}{confer}
\cite{Jut97,Sch01}.

\subsection{From NSW to DPW}

We now present our construction.
Let $\cS = \zug{\Sigma, S, \delta, s_0, \alpha}$ be an NSW where $\alpha = \{
\zug{R_1,G_1}, \ldots, \zug{R_k, G_k}\}$ and $|S|=n$. 
Denote $m=n(k+1)$. 
For the sake of the proof, we distinguish between the set of nodes
$V=[2m]$ of a tree and their names that range over $[m]$.
All important information (tree structure, label) can be associated
with the names and in practice the distinction between nodes and their
names is not needed.

\begin{defi}
A {\em compact Streett Safra tree} $t$ over $S$ is $\langle N, M, 1, p,
  l, h$, $e, f\rangle$ where the components of $t$ are as follows.
\begin{enumerate}[$\bullet$]
\item
  $N \subseteq V$ is a set of nodes.
\item
  $M: N \rightarrow [m]$ is the naming function.
\item
  $1 \in N$ such that $M(1)=1$ is the root node.
\item
  $p:N \rightarrow N$ is the parent function.
\item
  $l: N \rightarrow 2^S$ is a labeling of the nodes with subsets of
  $S$. 
  The label of every node is equal to the union of the labels of its
  sons. 
  The labels of two siblings are disjoint. 
\item
  $h: N \rightarrow 2^{[k]}$ annotates every node with a set of
  indices from $[k]$.
  The root is annotated by $[k]$.
  The annotation of every node is contained in that of its parent and
  it misses at most one element from the annotation of the parent.
  Every node that is not a leaf has at least one son with strictly
  smaller annotation. 
\item
  $e,f \in [m+1]$ are used to define the parity acceptance condition.
\end{enumerate}
\end{defi}

Notice that we give up the ``older than'' relation and replace the
sets $E$ and $F$ by numbers $e$ and $f$.
The naming $M$ is a bijection from $N$ to $[|N|]$. 
That is, the names of nodes \shorten{in $N$}{} are consecutive
starting from the root\shorten{, which is named $1$}{}. 

The following claim is proved much like the similar proof for Streett
Safra trees.
\begin{claim}  \Clml{compact streett safra trees}%\hfill
  The number of compact Streett Safra trees over $S$ is not more
  than \[\dpwsstates.\]
\end{claim}
\newcommand{\proofofcompactstreettsafratrees}{
\myproof{5}{
  Just like Streett Safra trees there are at most $m$ nodes.
  We use only the names of the nodes.
  The parent of the node has a smaller name.
  Thus, the parenthood relation can be represented by a sequence of
  (at most) $m-1$ pointers, where the $i$th pointer is pointing to a
  value in $1,\ldots, i{-}1$.
  It follows that there are at most $(m{-}1)!$ such trees.
  As the labels of the leaves form a partition of the set of states
  $S$ there are at most $n$ leaves.
  We add the function $l:S \rightarrow [n]$ that associates a state
  with the unique leaf to which it belongs. 
  Setting $l(s)=i$ means that $s$ belongs to the $i$th leaf.
  We can represent the annotation $h$ by annotating every edge by the
  value $j\in [k]$ such that $j$ is in the annotation of the parent
  and not in the annotation of the son. 
  If no such $j$ exists then we annotate the edge by $0$.
  The edge annotation is represented by a function $h:[m] \rightarrow
  [0,\ldots, k]$.
  In order to define the acceptance condition (see below) we need to
  know the value of $e$ in case $e\leq f$ and the value of $f$ in case
  $f<e$. 
  Thus, we need $2m$ possible values.
  
  It follows that the number of compact Streett Safra tress is at most
  $2m \cdot (m{-}1)! \cdot n^n \cdot (k+1)^m = 2m! \cdot n^n \cdot
  (k+1)^m = \dpwsstates$.
} % end of proof 5
}
\proofofcompactstreettsafratrees

\noindent
We construct the DPW $\cD$ equivalent to $\cS$.
Let $\cD = \zug{\Sigma, D, \rho, d_0, \alpha'}$ where the components of
$\cD$ are as follows.

\begin{enumerate}[$\bullet$]
\item
  $D$ is the set of compact Streett Safra trees over $\cS$.
\item
  $d_0$ is the tree with a single node $1$ labeled 
  $\{s_0\}$, named $1$, and annotated $[k]$. We set $e=2$ and
  $f=1$.
\item
  The parity acceptance condition $\alpha'{=}\zug{F_0,\ldots, F_{2m-1}}$
  is defined as follows.
  \begin{enumerate}[-]
  \item
    $F_0 = \{ d\in D ~|~ f=1 \mbox{ and } e>1 \}$
  \item
    $F_{2i+1} = \{ d\in D ~|~ e=i+2 \mbox{ and } f \geq e\}$
  \item
    $F_{2i+2} = \{ d\in D ~|~ f=i+2 \mbox{ and } e > f\}$
  \end{enumerate}
  As before, the case where $e=1$ is a rejecting sink state.
\item
  For every tree $d\in D$ and letter $\sigma \in \Sigma$ the
  transition $d'=\rho(d,\sigma)$ is the result of the following
  (recursive) transformation applied on $d$ starting from the root.
  Before we start, we set $e$ and $f$ to $m+1$ and replace the label
  of every node $v$ by $\delta(l(v),\sigma)$.

  \begin{enumerate}[(1)]
    \item
      If $v$ is a leaf such that $h(v)=\emptyset$ stop.
    \item
      If $v$ is a leaf such that $h(v)\neq \emptyset$, add to $v$
      a new son $v'$. 
      Set $l(v')=l(v)$, $h(v')=h(v)-\{max(h(v))\}$, and set $M(v')$
      to the minimal value greater than all used names.
      We may use temporarily names out of the range $[m]$.
      \label{item two 2}
    \item
      Let $v_1,\ldots, v_l$ be the sons of $v$ (ordered according
      to their names) and let $j_1,\ldots, j_l$ be the
      indices such that $j_i=max((h(v)\cup \{0\})-h(v_i))$ (note that
      $|h(v)-h(v_i)|\leq 1$; in case that $h(v)=h(v_i)$ we have
      $j_i=0$).
      Call recursively the entire procedure on $v_1,\ldots, v_l$
%      (including sons created in step \ref{item two 2} above).
      (including sons created in step 2 above).
      
      For every son $v_i$ and every state $s \in l(v_i)$ do the following. 
      \begin{enumerate}[(a)]
      \item
	If $s\in R_{j_i}$,
	remove $s$ from the label of $v_i$ and all its descendants.
	Add a new son $v'$ to $v$.
	Set $l(v') = \{s\}$, $h(v') = h(v)- \{max(\{0\} \cup (h(v)\cap
	\{1,\ldots, j_i-1\}))\}$, and set $M(v')$ to the minimal
	value larger than all used names.
	\label{moving good states}
      \item
	If $s\in G_{j_i}$,
	remove $s$ from the label of $v_i$ and all its descendants.
	Add a new son $v'$ to $v$.
	Set $l(v') = \{s\}$, $h(v') = h(v)-\{j_i\}$, and set
	$M(v')$ to the minimal value larger than all used names.
	\label{moving bad states}
      \end{enumerate}
    \item
      If a state $s$ appears in $l(v_i)$ and $l(v_{i'})$ and
      $j_i < j_{i'}$ then remove $s$ from the label of $v_{i'}$ and
      all its descendants.
    \item
      If a state $s$ appears in $l(v_i)$ and $l(v_{i'})$,  
      $j_i=j_{i'}$, and $M(v_i)<M(v_{i'})$ then remove $s$ from the label
      of $v_{i'}$ and all its descendants.
    \item
      Remove sons with empty label. 
      Set $e$ to the minimum of its previous value and the minimal
      name of a removed descendant. 
    \item
      \label{step in construction 2}
      If all sons are annotated by $h(v)$ remove all sons and all
      their descendants.
      Set $e$ to the minimum of its previous value and the minimal
      name of a removed descendant.
      Set $f$ to the minimum of its previous value and the name of
      $v$.
  \end{enumerate}
  Let $Z$ denote the set of nodes removed during this recursive
  procedure. 
  For every node $v$ let $rem(v)$ be $|\{v' \in Z ~|~ M(v') <
  M(v)\}|$.
  That is, we count how many nodes got removed during the
  recursive transformation and their name is smaller than the
  name of $v$. 
  For every node $v$ such that $l(v)\neq \emptyset$ we change the
  name of $v$ to $M(v)-rem(v)$. 
  The resulting names are consecutive again and in the range $[m]$.
\end{enumerate}

%\shorten{
\noindent
We show that the two automata are equivalent.
The proof is an adaptation of Safra's proof \cite{Saf92}.

\begin{thm}
  $L(\cD) = L(\cS)$.
  \Thml{language equivalence DPW NSW}
\end{thm}
%}{}
}
\newcommand{\nswtodpwa}{

\myproof{6}{
  Consider $w\in L(\cS)$.
  We have to show $w\in L(\cD)$.
  Let $r=s_0s_1\cdots$ be an accepting run of $\cS$ on $w$.
  Let $J \subseteq [k]$ be the maximal witness set for $r$.
  Let $r'=d_0d_1\cdots$ be the run of $\cD$ on $w$ and let $d_i =
  \zug{N_i,M_i,1,p_i,l_i,h_i,e_i,f_i}$.
  It is simple to see that for all $i\geq 0$ we have $s_i \in l_i(1)$
  and $e_i > 1$.
  Let $i_1$ be the location such that for all $i>i_1$ we have $s_i \in
  \inf(r)$. 
  That is, all states appearing after $i_1$ appear infinitely often in
  the run. 
  In particular, for all $i>i_1$ we have $s_i \notin G_j$ for all
  $j\notin J$.

%  If step~\ref{step in construction 2} is applied infinitely often to
  If step 7 is applied infinitely often to
  node 1 (equivalently, $f=1$ infinitely often, or during the
  application of transitions the descendants of 1 are all annotated by
  $[k]$) then $r'$ visits $F_0$ infinitely often.
%  Otherwise, from some point onwards in $r'$ we have step~\ref{step in
%    construction 2} is not applied to node $1$.
  Otherwise, from some point onwards in $r'$ we have step 7
    is not applied to node $1$.
  Let $i_2>i_1$ be this point.
  It follows that for all $i>i_2$ node $1$ is not a leaf.
  Then for all $i>i_2$ we have $s_i$ appears in the label of some
  son of $1$.
  This son can be changed a finite number of times. 
  The annotation of the edge to the son containing $r$ can only decrease. 
  If the edge is annotated by some $j \in J$ then $r$ eventually
  visits again $R_j$ and $r$ is migrated to some son annotated by
  $j'<j$.
  If the edge is annotated by some $j \notin J$ then $r$ never visits
  $G_j$ again and the only way to migrate to a different son is if $r$
  somehow appears again in a different son with smaller annotation, or
  if $r$ appears again in a different son with smaller name.
  Obviously, this can happen a finite number of times and eventually
  $r$ stays in the same son of $1$.
  The edge to this son is either annotated by $0$ or by some $j_1
  \notin J$.
  Formally, let $i_3>i_2$ be such that for all $i>i_3$ we have $s_i$
  appears in $l_i(v_1)$ and $v_1$ is a son of $1$.
  We know that for all $i>i_3$ we have $J \subseteq h_i(v_1)$.
  The name $M(v_1)$ may decrease finitely often until it is constant.
  Let $i_4> i_3$ be such that for all $i>i_4$ we have $a_1=M_i(v_1)$.
  As $M_i(v_1)=a_1$ for all $i>i_4$ it follows that $e_i>a_1$ for all
  $i>i_4$. 

%  If step~\ref{step in construction 2} is applied to node $v_1$
  If step 7 is applied to node $v_1$
  infinitely often then we are done.
  Otherwise, we construct by induction a sequence $1,v_1, \ldots, v_o$
  such that eventually $v_1,\ldots, v_o$
  do not change their names and $r$ appears in the label of all of
  them.
  Furthermore, we have $J \subseteq h(v_o)$ (which implies that
  $J\subseteq h(v_{o'})$ for all $o \in [1..o]$).
  As the number of active nodes in a tree is bounded by $m$ we can
  repeat the process only finitely often.
  Hence, $w$ is accepted by $\cD$.

  In the other direction, consider $w\in L(\cD)$. 
  Let $r' = d_0d_1\cdots$ be the accepting run of $\cD$ on $w$ where
  $d_i=\zug{N_i,M_i,1,p_i,l_i,h_i,e_i,f_i}$.
  Let $F_{2a}$ be the minimal set to be visited infinitely often.
  It follows that eventually always $e_i > a+1$ and infinitely often
  $f_i=a+1$.
  
  We write in short {\em avoids $G_{_{\overline{J}}}$} instead of
  avoids $G_{j}$ for every $j\notin J$ and {\em visits $R_{_J}$}
  instead of visits $R_j$ for every $j\in J$.
  We first prove two claims. 
%  The first, showing that all the states of $\cS$ that appear in
%  labels of nodes of a state of $\cD$ are reachable from the initial
%  state of $\cS$. 
%  The second, proves that if for some $2j$ the set $F_{2j}$ is
%  visited in $d_i$ and $d_{i'}$ and no visit to $F_{j'}$ for $j'<2j$
%  occurs between $i$ and $i'$, then there exists a node $v$ such that 
%  $M_a(v)=j+1$ for all $i \leq a \leq i'$ and for every state $s$ in
%  $l_{i'}(v)$ we  find a run segment of $\cS$ that starts in some
%  state in $l_i(v)$, avoids $G_j$ for all $j\notin h_i(v)$, visits
%  $R_j$ for all $j\in h_i(v)$, and ends in $s$.

  \begin{claim}
    For every $i\in \mathbb{N}$, $v\in N_i$, and every state $s\in
    l_i(v)$ we have $s$ is reachable from $s_0$ reading $w[0,i-1]$.
    \Clml{reachability dpw nsw}
  \end{claim}

  \myproof{6.1}{ 
    We prove the claim simultaneously for all $v \in N_i$ by induction
    on $i$. 
    Clearly, it holds for $i=0$.
    Suppose that it holds for $i$.
    As $l_{i+1}(v) \subseteq \delta(l_i(v'),w_i)$ for some $v' \in N_i$  
    it follows that every state in $l_{i+1}(v)$ is reachable from $s_0$
    reading $w[0,i]$.
  } % end of proof 6.1

  The following claim shows that if some node $v$ is colored green
  twice, and it is not removed and does not change its name between
  the two greens, then all the runs followed by $v$ visit $R_{_J}$, 
  where $J$ is the annotation of $v$.
  We essentially prove that all states that are followed by a son $v'$
  of $v$ annotated by $j \in J$ are endpoints for runs that already
  visited $R_{j'}$ for all $j' \in J$ such that $j'>j$.
  When $v$ is colored green the second time, all states followed by
  $v$ are in sons annotated $0$. This means that all $R_J$ is visited.

  \begin{claim}
    Consider $i,i' \in \mathbb{N}$ such that $i<i'$,
    $d_i,d_{i'} \in F_{2a}$
    for some $a$, and for all $a'\leq 2a$ and for every $o\in [i..i']$
    we have $d_{o} \notin F_{a'}$.
    Then there exists a node $v$ such that $M_{o}(v)=a+1$ for every
    $o \in [i..i']$ and every state $s$ in $l_{i'}(v)$ is reachable from
    some state in $l_i(v)$ reading $w[i,i'-1]$ with a run that avoids
    $G_{_{\overline{J}}}$ and visits $R_{_J}$.
    \Clml{visit to F dpw nsw}
  \end{claim}

  \myproof{6.2}{
    As $d_i \in F_{2a}$, there exists some node $v$ such that
    $M_i(v)=a+1$.
    By assumption, for every $a'<2a$ the set $F_{a'}$ is not visited
    between $i$ and $i'$. 
    Hence, for every node $v'$ such that $M_i(v') \leq a+1$ and for
    every $o \in [i..i']$ we have that $M_o(v')=M_i(v')$.
    That is, between $i$ and $i'$ all nodes whose name is at most
    $a+1$ do not change their names.
    In particular, for every $o\in [i..i']$ we have
    $M_o(v)=a+1$.
    In addition, there exists $J\subseteq [k]$ such that for every 
    $o\in [i..i']$ we have $h_o(v)=J$.

    We find a run followed by node $v$ between $i$ and
    $i'$ that avoids $G_{_{\overline{J}}}$ and visits $R_{_J}$.

    We first show that all runs followed by $v$ avoid
    $G_{_{\overline{J}}}$.
    One of the invariants maintained by the transition is that if a
    node $v$ is annotated by a set $J$ then it cannot be labeled by
    states in $G_j$ for $j\notin J$.
    Formally, suppose that for some $o \in [i..i']$ there exists $s\in
    l_o(v)$ such that $s\in G_j$ for some $j\notin J$.
    Let $v'$ be the youngest (according to the parenthood relation)
    ancestor of $v$ such that $j \in h_o(v')$ and let $v''$ be the son
    of $v'$ that is an ancestor of $v$ (it may be $v$ itself).
    It follows that the edge from $v'$ to $v''$ is labeled by $j$.
%    Then, when applying step~\ref{moving bad states} on the
    Then, when applying step 3b on the
    transformation from $d_{o-1}$ to $d_o$ the state $s$ would have
    been moved from $v''$ to some other son of $v'$.
    
    We show now that for every $o \in [i..i']$ and every $s \in l_o(v)$
    such that $s$ appears in a son of $v$ whose edge is annotated $j
    \in J$ there exists a run starting in some state in $l_i(v)$,
    visiting $R_{[(j+1)..k] \cap J}$, reading $w[i,o-1]$, and ending
    in $s$.
    We prove this by induction on $o$.
    The first thing in the transformation from $d_i$ to $d_{i+1}$ is
    to put all the elements in $l_{i+1}(v)$ in a son labeled by
    $max(J)$. Clearly, this satisfies our requirement.
    Suppose that it is true for $o$ and prove for $o+1$.
    Consider a state $s$ appearing in $l_{o+1}(v)$ in a son $v'$ such
    that the edge $(v,v')$ is annotated by $j$.
    If there is a predecessor of $s$ in the same son in $d_o$ then the
    claim follows (this covers the case where the same state appears
    in a node with smaller annotation or in a node with same
    annotation but smaller name).
%    Otherwise, $s$ appears in a son created by step~\ref{moving good
%    states}.
    Otherwise, $s$ appears in a son created by step 3a.
    It follows that there is some predecessor $s'$ of $s$ in a son
    $v''$ of $v$ in $d_o$ such that $(v,v'')$ is annotated by the
    minimal $j'>j$ such that $j' \in J$.
    Then, by induction there exists a run that ends in $s'$ and visits
    $R_{[(j'+1)..k] \cap J}$.
    In addition $s$ is in $R_{j'}$.
    The claim follows.

    As during the transformation from $d_{i'-1}$ to $d_{i'}$ all the
    states $s\in l_{i'}(v)$ are found in sons whose edge is annotated
    by $0$ we conclude that every state $s \in l_{i'}(v)$ is reachable
    along a run that visits $R_{_J}$.
  } % end of proof 6.2

  We find a witness set $J\subseteq [k]$ and 
  construct an infinite tree with finite branching degree.
  The root of the tree corresponds to the initial state of $\cS$.
  Every node in the tree is labeled by some state of $\cS$ and a time
  stamp $i$. 
  An edge between the nodes labeled $(s,i)$ and $(t,i')$ corresponds to
  a run starting in $s$, ending in $t$, reading $w[i,i'-1]$, avoiding
  $G_{_{\overline{J}}}$, and visiting $R_{_J}$.
  From K\"onig's lemma this tree contains an infinite branch. 
  The composition of all the run segments in this infinite branch is
  an infinite accepting run of $\cS$ on $w$ according to witness set $J$. 

%  We are now ready to build the tree $t$. 
  Let $(s_0,0)$ label the root of $T$. 
  Let $i_0$ be the minimal location such that for all $a'<2a$ the set
  $F_{a'}$ is not visited after $i_0$.
  Let $v$ be the node such that for all $i>i_0$ we have $M_{i}(v)=a+1$.
  Let $J \subseteq [k]$ be such that for all $i>i_0$ we have $h_{i}(v)=J$.
  Let $i_1$ be the minimal location such that $i_1> i_0$ and
%  $f_{i_1}=a+1$ (that is step~\ref{step in construction 2} was applied
  $f_{i_1}=a+1$ (that is step 7 was applied
  to $v$). 
  For every state $s$ in $l_{i_1}(v)$ we add a node to $T$, label it
  by $(s,i_1)$ and connect it to the root. 
  We extend the tree by induction.
  We have a tree with leaves labeled by the states in $l_{i_o}(v)$
%  stamped by time $i_o$, and $f_{i_o}=a+1$ (step~\ref{step in construction 2}
  stamped by time $i_o$, and $f_{i_o}=a+1$ (step 7
  was applied to $v$).
  That is, for every state $s$ in $l_{i_o}(v)$ there exists a leaf
  labeled $(s,i_o)$.
  We know that $F_{2a}$ is visited infinitely often. 
  Hence, there exists a minimal $i_{o+1} > i_o$ such that
%  $f_{i_{o+1}}=a+1$ (step~\ref{step in construction 2} is applied to
  $f_{i_{o+1}}=a+1$ (step 7 is applied to
  $v$).
  For every state $s'$ in $l_{i_{o+1}}(v)$ we add a node to the tree and 
  label it $(s',i_{o+1})$. 
  From \Clmr{visit to F dpw nsw} there exists a state $s$ in $l_{i_o}(v)$
  such that $s'$ is reachable from $s$ reading $w[i_o,i_{o+1}-1]$ with a run
  that avoids $G_{_{\overline{J}}}$ and visits $R_{_J}$. 
  We connect $(s',i_{o+1})$ to $(s,i_o)$.

  From \Clmr{reachability dpw nsw} it follows that every edge
  $(s_0,0),(s',i_1)$ corresponds to some run starting in $s_0$, ending
  in $s'$, and reading $w[0,i_1-1]$. 
  From \Clmr{visit to F dpw nsw}, every other edge in the tree
  $(s,i_o),(s',i_{o+1})$ corresponds to some run starting in $s$, ending in
  $s'$, reading $w[i_o,i_{o+1}-1]$, avoiding $G_{_{\overline{J}}}$, and
  visiting $R_{_J}$.
  From K\"onig's lemma there exists an infinite branch in the tree.
  This infinite branch corresponds to an accepting run of $\cS$
  on $w$.
} % end of proof 6  
}

\nswtodpw
\nswtodpwa

\begin{thm}%\hfill
  For every NSW $\cS$ with $n$ states and index $k$ there exists a DPW
  $\cD$ with $\dpwsstates$ states and index $2n(k+1)$ such that
  $L(\cD)=L(\cS)$. \qed
  \Thml{dpw nsw complexity}
\end{thm}

\noindent
As before, when compared to Safra's construction, we reduce the number
of states and get a parity automaton.
The advantages are similar to those described in \Secr{dpw}.

\shorten{
\section{Conclusions and Future Work}
\Secl{future}

We improved both of Safra's determinization constructions. 
In both cases, we reduce the number of states and more important
construct directly a parity automaton.
In the case of NBW we reduce the maximal number of states from
$\safrastates$ to $2n^nn!$. 
In the case of NSW we reduce the maximal number of states from 
$\safrasstates$ to $\dpwsstates$. 
The fact that our automata are parity automata makes them easier to
use `down the line'.
The algorithms for solving parity games are much simpler than those
that solve Rabin games.
In particular, Rabin games are NP-complete in the Rabin index while
parity games are known to be in NP$\cap$co-NP.
The complement of a DPW is again a DPW.
In contrast, the complement of a DRW is a DSW.
In order to get back to Rabin (or parity) one has to multiply the
number of states by $k^2k!$, where $k$ is the number of Rabin pairs of
the automaton. 
Our upper bound improves the best known upper bound in
numerous applications, such as solving games, complementation of tree
automata, emptiness of alternating tree automata, satisfiability of
$\mu$-calculus with backward modalities and CTL$^*$.
In particular, in the recent emptiness algorithm for 
alternating parity tree automata \cite{KV05c} the upper bound is
%reduced from $(12)^{n^2}n^{4n^2}n!$ to $n^{4n^2}$.
reduced from $(12)^{n^2}n^{4n^2+2n}(n!)^n$ to $(2n^nn!)^{2n}$.

There are lower bounds for both determinization constructions.
For an NBW with $n$ states the best possible DPW has at least $n!$
states \cite{Mic88}. 
For an NSW with $n$ states and $k$ Streett pairs the best possible DPW
has at least $(\Omega(nk))^n$ states \cite{Yan06}.
We have gotten closer to this lower bound however there is still a
large gap between the lower bound and the upper bound.
We are not aware on similar lower bounds on the index of the resulting
automata. 
As DPW[k+1] recognize more languages than DPW[k] \cite{Wag79} and NBW
recognize all $\omega$-regular languages we cannot hope for a
determinization construction with constant index.
The language $L_k = \{ w \in [1..k]^\omega ~|~ min(inf(w)) \mbox{ is
even}\}$ is in DPW[k] but not in DPW[k-1].
It is simple to construct an NBW with $k$ states recognizing $L_k$.
This suggests that a determinization of NBW with $k$ states may result
in DPW with $k$ priorities.
It is an interesting question whether the $2k$ priorities produced by
our construction are indeed necessary.
A similar question arises for NSW.
}{}

\section*{Acknowledgment}
I thank T.A. Henzinger for fruitful discussions,
O. Kupferman and M.Y. Vardi for discussions on Safra's construction
and comments on an earlier version,
Y. Lustig for comments on an earlier version and for 
tightening the analysis of the number of states, and the
referees for comments and suggesting the lower bound on
the index of NBW.

%% in general the use of bibtex is encouraged

\bibliographystyle{alpha}
\bibliography{/cygdrive/d/Bib/ok}
\end{document}